\documentclass[12pt]{article} \usepackage{amssymb,latexsym}
\usepackage{amsmath}
 
\columnwidth\textwidth

\begin{document}

\newtheorem{df}{Definition} \newtheorem{thm}{Theorem} \newtheorem{lem}{Lemma}
\newtheorem{conj}{Conjecture} \newtheorem{assump}{Assumption}
 
\begin{titlepage}
 
\noindent
 
\begin{center} {\LARGE Quantum models of classical systems} \vspace{1cm}

P. H\'{a}j\'{\i}\v{c}ek\\ Institute for Theoretical Physics \\ University of
Bern \\ Sidlerstrasse 5, CH-3012 Bern, Switzerland \\ hajicek@itp.unibe.ch

\vspace{1cm}

December 2014 \\ \vspace{1cm}
 
 
\vspace*{2cm}
 
\nopagebreak[4]
 
\begin{abstract} Quantum statistical methods that are commonly used for the
derivation of classical thermodynamic properties are extended to classical
mechanical properties. The usual assumption that every real motion of a
classical mechanical system is represented by a sharp trajectory is not
testable and is replaced by a class of fuzzy models, the so-called maximum
entropy (ME) packets. The fuzzier are the compared classical and quantum ME
packets, the better seems to be the match between their dynamical
trajectories. Classical and quantum models of a stiff rod will be constructed
to illustrate the resulting unified quantum theory of thermodynamic and
mechanical properties.
\end{abstract}

\end{center}

\end{titlepage}

\section{Introduction} There are some features of the classical world that
seem to be incompatible with quantum mechanics:
\begin{description}
\item[Realism] Properties such as position and momentum can be ascribed to a
chair, say, independently of whether they are observed or not.
\item[Sharp Trajectories] By a common interpretation of classical mechanics,
the real chair is even at a sharp point of its phase space at each
time. Attempts to model this property by a quantum state with minimum
uncertainty leads to coherent states that are pure.
\item[No Superpositions] The chair is never observed in a linear superposition
of being, e.g., simultaneously in the kitchen as well as in the
bedroom. However, pure states in quantum mechanics can be superposed in this
manner.
\item[Robustness] Measurement of every classical observable can be done in
such a way that the state of the observed system is arbitrarily weakly
disturbed. However, pure quantum states are not disturbed only by measurements
of very few very special observables.
\end{description} Thus, attempts to solve the problem of Sharp Trajectories
aggravate problems of Robustness and of No Superpositions.

There is a vast literature about the problems. Let us list examples of the
most popular ideas: macroscopic systems do not obey quantum mechanics
\cite{leggett}; quantum decoherence theory \cite{schloss}; only coarse-grained
operators represent classical measurements \cite{kampen}; Coleman-Hepp theory
\cite{hepp}; dynamical collapse theory \cite{pearle,ghirardi}. The list is
incomplete.

Our theory is different. It rejects sharp trajectories and seeks quantum
mechanical derivation of classical properties possessed by fuzzy mechanical
states. The present paper is a short review of \cite{hajicek,entropy} as well
as of some new results.

\section{Hypothesis of high entropy states} To motivate our approach, let us
briefly recapitulate some ideas of statistical thermodynamics. Consider
rarefied equilibrium gas in a vessel. There is a classical model ${\mathcal
S}_c$ of this gas offered by phenomenological thermodynamics, called ``ideal
gas'', and the properties of ${\mathcal S}_c$ are examples of classical
properties. They are described by thermodynamic quantities such as internal
energy $E$, volume $\Omega$, pressure, entropy, temperature, specific heats,
etc.

To obtain the values of such quantities from quantum mechanics, we need a
quantum model ${\mathcal S}_q$ of the gas. As ${\mathcal S}_q$, we can choose
a system of $N$ spin-zero point particles, each with mass $\mu$, in a deep
potential well of volume $\Omega$ with Hamiltonian
$$
{\mathsf H} = \sum_{k=1}^N\frac{|\vec{\mathsf p}_k|^2}{2\mu}\ ,
$$
where $\vec{\mathsf p}_k$ is the momentum of $k$-th particle in the rest
system of $\Omega$. ${\mathsf H}$ is then the operator of the internal energy
and the classical internal $E$ energy is an average of ${\mathsf H}$.

The most important assumption of the quantum model is the choice of state. It
is the state that maximises the (von Neumann) entropy for fixed value $E$ of
the average of the internal energy. It is called ``Gibbs state''. All
properties of the ${\mathcal S}_c$ can then be calculated from ${\mathcal
S}_q$ as properties of the Gibbs states.

The main (heuristic) principle of our theory is a generalisation of this idea
to all classical properties, including the mechanical ones. Thus, we state the
following hypothesis:
\begin{assump}\label{highS} Let a real system ${\mathcal S}$ have a classical
model ${\mathcal S}_c$. Then, there is a quantum model ${\mathcal S}_q$ of
${\mathcal S}$ such that all properties of ${\mathcal S}_c$ are selected
properties of some high-entropy states of ${\mathcal S}_q$.
\end{assump} An important reason for accepting this hypothesis is that it
suggests ways in which all four problems mentioned in the introduction can be
solved. Indeed, the Realism Problem could be approached as follows. Our theory
of objective properties of quantum systems \cite{PHJT,hajicek5} justifies the
assumption that quantum states are objective. If classical properties are
properties of some states of the quantum model, they will also be
objective. The No-Superposition Problem is based of some properties of pure
quantum states. But high-entropy states are not pure: they cannot be
superposed. As for the robustness problem, we can use the fact that very many
quantum states correspond to a single classical state. Even if quantum states
may be disturbed by observation, the corresponding classical states need not
be. Finally, there is no Sharp-Trajectories Problem for thermodynamics. All
these points are just suggestions and must be more carefully studied on some
mathematically well-defined models.

Assumption \ref{highS} might work for thermodynamics, but what could be the
high-entropy states for Newtonian mechanics?

\section{Classical ME packets} Let us consider the classical mechanical model
${\mathcal S}_c$ defined as a system with a single degree of freedom and
Hamiltonian
\begin{equation}\label{comhamc} H = \frac{p^2}{2\mu} + V(q)\ .
\end{equation} The classical equations of motion are
\begin{equation}\label{motionc} \dot{q} = \frac{p}{\mu}\ ,\quad \dot{p} =
-\frac{dV}{dq}
\end{equation} and their solution is a sharp trajectory
$$
q = q(t)\ ,\quad p = p(t)
$$
for every initial values $q(0)$ and $p(0)$.

Let us choose the corresponding quantum model ${\mathcal S}_q$ to be a system
of one degree of freedom with position operator ${\mathsf q}$, momentum
operator ${\mathsf p}$ and spin 0. Let the Hamiltonian be
\begin{equation}\label{comham} {\mathsf H} = \frac{{\mathsf p}^2}{2\mu} +
V({\mathsf q})\ .
\end{equation} The Heisenberg equations of motion are
\begin{equation}\label{motionq} \dot{\mathsf q} = \frac{{\mathsf p}}{\mu}\
,\quad \dot{\mathsf p} = -\frac{dV}{d{\mathsf q}}\ .
\end{equation} Then the time dependence of position and momentum averages $Q =
\langle {\mathsf q} \rangle$ and $P = \langle {\mathsf p} \rangle$ in a state
$|\psi\rangle$ is
$$
\dot{Q} = \frac{P}{\mu}\ ,\quad \dot{P} = -\left\langle \frac{dV}{d{\mathsf
q}}\right\rangle\ .
$$
To evaluate the right-hand side of the second equation, let us expand the
potential function in powers of ${\mathsf q} - Q$:
$$
V({\mathsf q}) = V(Q) + ({\mathsf q} - Q)\frac{dV}{dQ} + \frac{1}{2}({\mathsf
q} - Q)^2 \frac{d^2V}{dQ^2} + \ldots
$$
so that
$$
\frac{dV}{d{\mathsf q}} = \frac{dV}{dQ} + ({\mathsf q} - Q)\frac{d^2V}{dQ^2} +
\frac{1}{2}({\mathsf q} - Q)^2 \frac{d^3V}{dQ^3} + \ldots\ .
$$
If we take the average of the last equation and use relations $\langle
({\mathsf q} - Q)\rangle = 0$ and $\langle ({\mathsf q} - Q)^2\rangle = \Delta
Q^2$, where $\Delta Q$ is the variance of ${\mathsf q}$ in state
$|\psi\rangle$, we obtain
$$
\left\langle \frac{dV}{d{\mathsf q}}\right\rangle = \frac{dV}{dQ} +
\frac{1}{2}\Delta Q^2 \frac{d^3V}{dQ^3} + \ldots\ .
$$

Let us assume that coordinate $q$ and momentum $p$ of ${\mathcal S}_c$ are
obtained from the quantum model by formulas
$$
q = Q\ ,\quad p = P\ .
$$
Then, already for potentials of the third order, the quantum equations of
motion for averages deviate from classical equation of motion for sharp
trajectories. This deviation would be negligible for small $\Delta Q$, that
is, the spread of the wave packet $|\psi\rangle$ over the space must be as
small as possible. However, if the variance $\Delta P$ is large, $\Delta Q$
will quickly increase with time. This implies that the minimum-uncertainty
wave packets may give the best approximation to classical sharp trajectories.

Let us stop here and ask: what is the reason for trying to get sharp
trajectories from quantum mechanics? Clearly, it is the popularity of the
specific form of classical realism mentioned in the Introduction: a {\em real}
mechanical system possesses a sharp position and momentum at any instant of
time. Let us call this assumption {\bf Sharp Trajectory Hypothesis}
(STH). There are many tacitly assumed consequences of STH, for example that a
probability distributions on phase space is only an expression of insufficient
knowledge of the real state.

However, there is {\em no evidence} supporting STH: indeed, as yet, any real
observation of macroscopic bodies has been compatible with
\begin{equation}\label{NSTH} 2\Delta Q\Delta P \gg \hbar\ ,
\end{equation} where ``$\gg$'' represents many orders of magnitude. {\em This}
is well known but there can be {\em two} attitudes to Eq.\ (\ref{NSTH}):
\begin{enumerate}
\item With improving techniques, the left-hand side of Eq. (\ref{NSTH}) will
approach zero. This must be false if quantum mechanics holds true.
\item Sharp trajectory is just a handy model of a real, fuzzy, one. That is,
it lies within a tube associated with the fuzzy trajectory. But then, a more
realistic model of any Newtonian motion would be a probability distribution.
\end{enumerate} References \cite{PHJT,entropy} assume the second attitude. For
us, the most important consequence is that it is sufficient to approximate
{\em fuzzy} Newtonian trajectories by quantum mechanics, where fuzzy
trajectories are some probability distributions on the phase space of the
system. Such a theory of classical properties can do without pure states. Of
course, {\em this probability distribution is not completely knowable and
measurable: in any case, the sharp points do not exist} \cite{Exner,Born}. The
fact that the points of the phase space do exist mathematically and must be
used for mathematical description of a real state is only an unrealistic
feature of Newtonian mechanics.

Thus, instead of a sharp point of the phase space a distribution on the phase
space can be considered as the real state of a mechanical system. It is
determined by preparation similarly as in quantum theory. In this way, we
preserve the realism (for more details on realism, see \cite{hajicek5}) as
such but change the form of it as expressed by STH.

This opens the problem to application of Bayesian methods, see, e.g.,
\cite{Jaynes}. These methods recommend maximising entropy in the cases of
missing knowledge. Let us define a fuzzy state called {\em maximum-entropy
packet} (ME packet) as a phase-space distribution maximising entropy for given
averages and variances of mechanical state coordinates.

More precisely, for the classical model ${\mathcal S}_c$, we consider the
states described by distribution function $\rho(q,p)$ on the phase space
spanned by $q$ and $p$. The function $\rho(q,p)$ is dimensionless and
normalised by
$$
\int\frac{dq\,dp}{v}\,\rho = 1\ ,
$$
where $v$ is an auxiliary phase-space volume to make $\rho$ dimensionless. The
entropy of $\rho(q,p)$ can be defined by
$$
S := -\int\frac{dq\,dp}{v}\,\rho \ln\rho\ .
$$
The value of entropy will depend on $v$ but most other results will
not. Classical mechanics does not offer any idea of how to fix $v$. We shall
get its value from quantum mechanics.

\begin{df}\label{dfold21} ME packet is the distribution function $\rho$ that
maximizes the entropy subject to the conditions:
\begin{equation}\label{21.4} \langle q\rangle = Q\ ,\quad \langle q^2\rangle =
\Delta Q^2 + Q^2\ ,
\end{equation} and
\begin{equation}\label{21.5} \langle p\rangle = P\ ,\quad \langle p^2\rangle =
\Delta P^2 + P^2\ ,
\end{equation} where $Q$, $P$, $\Delta Q$ and $\Delta P$ are given values.
\end{df} We have used the abbreviation
$$
\langle x\rangle = \int\frac{dq\,dp}{v}\,x\rho\ .
$$

The explicit form of $\rho$ can be found using the Lagrange-multiplier and
partition-function method \cite{hajicek}:
\begin{thm}\label{propold19} The distribution function of the classical ME
packet for a one-degree-of-freedom system with given averages and variances
$Q$, $\Delta Q$ of coordinate and $P$, $\Delta P$ of momentum, is
\begin{equation}\label{23.1} \rho[Q,P,\Delta Q,\Delta P](q,p) =
\left(\frac{v}{2\pi}\right)\frac{1}{\Delta Q\Delta
P}\exp\left[-\frac{(q-Q)^2}{2\Delta Q^2} -\frac{(p-P)^2}{2\Delta P^2}\right]\
.
\end{equation}
\end{thm}

In this way, to describe the mechanical degrees of freedom, we need twice as
many variables as the standard mechanics. The doubling of state coordinates is
due to the necessity to define a fuzzy distribution rather than a sharp
trajectory.

The model can be generalised to any number of degrees of freedom. Also, the ME
packet could be defined by different pairs of conjugate variables. It seems
plausible that our main results would then remain valid.

\section{Quantum ME packets}
\begin{df}\label{dfold22} State ${\mathsf T}$ of quantum model ${\mathcal
S}_q$ with one degree of freedom that maximizes von Neumann entropy
$$
S = -tr({\mathsf T}\ln{\mathsf T})
$$
under the conditions
$$
tr[{\mathsf T}{\mathsf q}] = Q\ ,\quad tr[{\mathsf T} {\mathsf q}^2] = Q^2 +
\Delta Q^2\ ,
$$
$$
tr[{\mathsf T}{\mathsf p}] = P\ ,\quad tr[{\mathsf T} {\mathsf p}^2] = P^2 +
\Delta P^2\ ,
$$
where $Q$, $P$, $\Delta Q$ and $\Delta P$ are given numbers, is called {\em
quantum ME packet}.
\end{df}

The following theorem can be proved by the method of Lagrange multipliers and
partition function, but the proof is non-trivial because of non-commuting
factors \cite{hajicek}:
\begin{thm}\label{propold20} The state operator of the ME packet of a
one-degree-of-freedom system with given averages and variances $Q$, $P$,
$\Delta Q$ and $\Delta P$ is
\begin{equation}\label{MEq} {\mathsf T}[Q,P,\Delta Q, \Delta P]=
\frac{2}{\sqrt{\nu^2-1}}\exp\left(-\frac{\nu}{2}\ln\frac{\nu+1}{\nu-1}\
{\mathsf K}\right)\ ,
\end{equation} where
$$
{\mathsf K} = \frac{({\mathsf q}-Q)^2}{2\Delta Q^2} + \frac{({\mathsf
p}-P)^2}{2\Delta P^2}
$$
and
\begin{equation}\label{Knu} \quad \nu = \frac{2\Delta P\Delta Q}{\hbar}\ .
\end{equation}
\end{thm} Generalisation to any number of degrees of freedom is easy. It is
amusing to observe how the forms of Eqs. (\ref{23.1}) and (\ref{MEq}) approach
each other in the limit $\Delta P\Delta Q \rightarrow \infty$. Indeed,
$$
\lim_{\nu\rightarrow\infty}\frac{\nu}{2}\ln\frac{\nu+1}{\nu-1} = 1\ .
$$

The entropy of state (\ref{MEq}) can be shown \cite{hajicek} to be an
increasing function of $\nu \in (1,\infty)$ diverging for $\nu \rightarrow
\infty$. For $\nu = 1$ (minimum quantum uncertainty), ${\mathsf T}$ is a pure
state with wave function
$$
\psi(q) = \left(\frac{1}{\pi} \frac{1}{2\Delta Q^2}\right)^{1/4}
\exp\left[-\frac{1}{4\Delta Q^2}(q-Q)^2 + \frac{iPq}{\hbar}\right]\ .
$$
This is just a Gaussian wave packet and the entropy is zero. Thus, quantum ME
packets are generalization of Gaussian wave packets.

\section{Comparing classical and quantum evolutions} Let us consider the time
evolution of the averages and variances for ME packet (\ref{23.1}) with
initial data $Q, P, \Delta Q, \Delta P$ at $t=0$ and let us define the {\em
classical trajectory} of the classical model ${\mathcal S}_c$ by the quadruple
$Q_c(t), P_c(t), \Delta Q_c(t), \Delta P_c(t)$. Let $Q_q(t), P_q(t), \Delta
Q_q(t), \Delta P_q(t)$ be an analogous trajectory for the quantum model
${\mathcal S}_q$ starting in state (\ref{MEq}). Each of the two trajectories
is described by four real functions so that they can be compared.

Let us first study the special case of at most quadratic potential:
$$
V(q) = V_0 + V_1 q + \frac{1}{2} V_2 q^2\ ,
$$
where $V_k$ are constants with suitable dimensions. If $V_1 = V_2 =0$, we have
a free particle, if $V_2 = 0$, it is a particle in a homogeneous force field
and if $V_2 \neq 0$, it is an harmonic or anti-harmonic oscillator. In these
cases, exact solutions can be found:
\begin{eqnarray*} Q_c(t) = Q_q(t) &=& f_0(t) + Q f_1(t) + P f_2(t)\ , \\
\Delta Q_c(t) =\Delta Q_q(t) &=& \sqrt{f_1^2(t)\Delta Q^2 + f_2^2(t)\Delta
P^2}\ , \\ P_c(t) = P_q(t) &=& g_0(t) + Q g_1(t) + P g_2(t)\ ,\\ \Delta P_c(t)
= \Delta P_q(t) &=& \sqrt{g_1^2(t)\Delta Q^2 + g_2^2(t)\Delta P^2}\ .
\end{eqnarray*} If $V_2 \neq 0$, the functions are
$$
f_0(t) = -\frac{V_1}{V_2}(1-\cos\omega t)\ ,\quad f_1(t) = \cos \omega t\
,\quad f_2(t) = \frac{1}{\xi}\sin\omega t\ ,
$$
$$
g_0(t) = -\xi\frac{V_1}{V_2}\sin\omega t\ ,\quad g_1(t) = -\xi\sin \omega t\
,\quad g_2(t) = \cos\omega t\ ,
$$
where
$$
\xi = \sqrt{\mu V_2}\ ,\quad \omega = \sqrt{\frac{V_2}{\mu}}\ .
$$
If $V_2 = 0$, we obtain
$$
f_0(t) = -\frac{V_1}{2\mu}t^2\ ,\quad f_1(t) = 1\ ,\quad f_2(t) =
\frac{t}{\mu}\ ,
$$
$$
g_0(t) = -V_1t\ ,\quad g_1(t) = 0\ ,\quad g_2(t) = 1\ .
$$

Hence, for at-most-quadratic potentials, classical and quantum trajectories
coincide. However, already for a third order potentials, there are non-trivial
quantum corrections. For example, if $V = V_3{\mathsf q}^3/6$, one can use
Heisenberg equations of motion (\ref{motionq}) to calculate the 9-th time
derivative of ${\mathsf p}$ where there is finally a term,
$$
\frac{d^9{\mathsf p}}{dt^9} = \ldots -\frac{125}{4}\frac{V_3^5}{\mu^4}{\mathsf
q}^6\ ,
$$
that has a quantum correction, namely
$$
\langle {\mathsf q}^6 \rangle_q = \langle {\mathsf q}^6 \rangle_c + 9\Delta
Q^6\nu^{-1} - 3 \Delta Q^6\nu^{-3}\ ,
$$
where $\nu$ is defined by Eq.\ (\ref{Knu}) and
$$
\langle {\mathsf q}^6 \rangle_c = Q^6 + 15 Q^4\Delta Q^2 + 45 Q^2 \Delta Q^4 +
15 \Delta Q^6\ .
$$
One can see from these equations that the quantum correction becomes
negligible in the limit $\Delta Q\rightarrow \infty$. In general, one can
show:
\begin{thm}\label{thmnu} For general polynomial potential, the classical and
quantum trajectories satisfy:
$$
\lim_{\Delta Q\rightarrow \infty,\Delta P\rightarrow \infty}\frac{Q_q(t) -
Q_c(t)}{Q_c(t)} = 0\ ,\quad \lim_{\Delta Q\rightarrow \infty,\Delta
P\rightarrow \infty}\frac{P_q(t) - P_c(t)}{P_c(t)} = 0\ ,
$$
and
$$
\lim_{\Delta Q\rightarrow \infty,\Delta P\rightarrow \infty}\frac{\Delta
Q_q(t) - \Delta Q_c(t)}{\Delta Q_c(t)} = 0\ ,\quad \lim_{\Delta Q\rightarrow
\infty,\Delta P\rightarrow \infty}\frac{\Delta P_q(t) - \Delta P_c(t)}{\Delta
P_c(t)} = 0\ ,
$$
for all $t$ for which the formulas make sense.
\end{thm} The proof is based on the calculation of averages of polynomials in
${\mathsf q}$ and ${\mathsf p}$ described in \cite{hajicek} and will be
published elsewhere.

Hence, the fuzzier the compared ME packets are, the better their dynamical
evolutions match each other. In other words, the classical limit for
mechanical degrees of freedom is
\begin{equation}\label{nularge} \Delta Q \rightarrow \infty\ , \quad \Delta P
\rightarrow \infty\ .
\end{equation} This seems to contradict the usual belief that the classical
physics is best approximated by minimum-uncertainty ($\nu = 1$) quantum
states. But the explanation of this paradox is simple. The two answers to the
question which states best approximate classical physics are different because
the questions asked are, in fact, different: the first one compares two fuzzy
states, the second one compares a quantum state with a sharp classical
trajectory.

The conjecture that the classical limit is given by equation (\ref{nularge})
seems to be plausible, but it has yet to be shown for non-polynomial potential
function, such as the Coulomb potential (Kepler orbits).

\section{Mechanical and thermostatic properties unified} We have tried to
prove that classical mechanical properties of an object can be obtained from
its quantum model as properties of high-entropy quantum states. However, this
also holds for classical thermostatic properties, such as internal energy,
temperature, entropy, specific heats etc., which suggests that the quantum
theory of classical properties can be based on a single principle. In the
present section, we try to show in more detail how such ``unified'' theory
could look like.

We use a very simple model so that its quantum equations are exactly solvable
and we can concentrate on conceptual questions.  As a real object ${\mathcal
S}$, consider a thin stiff rod of mass $M$ and length $L$ extended and moving
freely in one space dimension. Let its classical model ${\mathcal S}_c$ be a
one-dimensional continuum. Its (classical) state is determined by the values
of 5 quantities: internal energy $E_{\text{int}}$, average $X$ and variance
$\Delta X$ of its centre-of-mass coordinate as well as average $P$ and
variance $\Delta P$ of its total momentum.

Let the quantum model ${\mathcal S}_q$ be a chain of $N+1$ particles, each of
mass $\mu$. We denote the position operator of the $n$-th particle by
${\mathsf x}_n$ and that of its momentum by ${\mathsf p}_n$, $n = 1,\ldots
N+1$. Let the Hamiltonian of the quantum model be
$$
{\mathsf H} = \frac{1}{2\mu}\sum_{n=1}^{N+1} {\mathsf p}_n^2 +
\frac{\kappa^2}{2}\sum_{n=2}^{N+1} ({\mathsf x}_n - {\mathsf x}_{n-1} -
\xi)^2\ .
$$
The potential represents nearest-neighbour elastic forces, $\kappa$ being the
oscillator strength and $\xi$ the equilibrium inter-particle distance. The
algebra of observables of ${\mathcal S}_q$ is generated by ${\mathsf x}_n$ and
${\mathsf p}_n$, a set of $2N + 2$ operators. We assume that $N \approx
10^{23}$. This implies that the quantum state contains much more information
than the classical one.

A linear (in fact, Fourier) transformation of variables ${\mathsf x}_n$ and
${\mathsf p}_n$ to normal modes ${\mathsf u}_n$ and ${\mathsf q}_n$
diagonalizes the Hamiltonian \cite{PHJT,entropy}. Moreover, it becomes the sum
of the total momentum part and the internal energy part ${\mathsf
E}_{\text{int}}$ (see \cite{PHJT,entropy}):
$$
{\mathsf H} = \frac{1}{2M}{\mathsf P}^2 + {\mathsf E}_{\text{int}}\ ,
$$
where $M = (N+1)\mu$ is the total mass of the chain, ${\mathsf P}$ its total
momentum,
$$
{\mathsf E}_{\text{int}} = \frac{1}{2\mu}\sum_{m=1}^{N}{\mathsf q}_m^2 +
\frac{\mu}{2}\sum_{m=1}^{N}\omega_m^2{\mathsf u}_m^2
$$
and $\omega_m$, $m = 1,\ldots,N$ are ``phonon'' frequencies:
$$
\omega_m = \frac{2\kappa}{\sqrt{\mu}}\,\sin\left(\frac{\pi}{2(N+1)}\,m\right)\
.
$$
The mechanical evolution thus decouples from the thermodynamics.

As the state of ${\mathcal S}_q$, the tensor product ${\mathsf
T}_{\text{therm}} \otimes {\mathsf T}_{\text{mech}}$ can, therefore, be
chosen. Next, we apply the unifying principle: ${\mathsf T}_{\text{therm}}$ is
the maximum entropy quantum state for a given value $E$ of the averages of
${\mathsf E}_{\text{int}}$ and ${\mathsf T}_{\text{mech}}$ is the maximum
entropy state for given averages of ${\mathsf X}$ and ${\mathsf P}$ and
variances $\Delta {\mathsf X}$ and $\Delta {\mathsf P}$. It then follows that
${\mathsf T}_{\text{therm}}$ is the Gibbs state and ${\mathsf
T}_{\text{mech}}$ is an ME packet.

The phonons of species $m$ form statistically independent subsystems. Hence,
the Gibbs state factorizes and the factors are Gibbs states ${\mathsf T}_m$ of
the species:
$$
{\mathsf T}_m = \sum_{r=0}^\infty |r\rangle {\mathrm P}^m_r \langle r|\ ,
$$
where $r$ is the number of phonons of species $m$,
$$
{\mathrm P}^m_r = \left(1 -
e^{-\lambda\hbar\omega_m}\right)e^{-\lambda\hbar\omega_m r}
$$
and $\lambda$ is the Lagrange multiplier of the variational problem for the
conditional maximum of entropy. The variational principle couples the value of
$\lambda$ with the value of the energy average $E$. As $\lambda$ is
interpreted as $1/kT$, $k$ being the Boltzmann constant and $T$ the
temperature, the relation between internal energy and temperature results. The
average number $\langle r\rangle$ of photons in state ${\mathsf T}_m$ is the
Bose distribution.

All properties of the classical model (such as the temperature and length of
the rod, its dynamical trajectory etc.) have been obtained, in a good
approximation, from the quantum one (\cite{PHJT,entropy}). For example the
length $L$ of the rod is the average of a natural rod-length operator
${\mathsf x}_{N+1} - {\mathsf x}_1$. The calculation in \cite{PHJT,entropy}
yields
$$
L = N\xi\ .
$$
This is independent of the parameter $E$ of the Gibbs state. Hence, the model
describes a rigid rod. The relative variances of the internal energy and
length are indirectly proportional to $N$.

\section{Conclusion and outlook} Our results suggest that there is a unified
theory for both thermostatic and mechanical properties. It is based on the
assumption that the states of quantum system that exhibit classical properties
are some states with high entropy.

The fuzzier are the compared mechanical states, the better seems to be the
match between classical and quantum mechanical trajectories. This would
confirm the feeling that quantum mechanics is more accurate and finer than
Newtonian mechanics. We hope to be able to prove this statement for
non-polynomial potentials later.

The paper suggests promising ideas of how all four conceptual problems can be
solved. More detailed models that would describe such solutions for some
simple cases ought to be constructed.

The project is in its beginnings. Only extremely simple models have been
studied. Also, a generalisation of the idea to classical electro- and
magnetostatic properties, as well as a generalisation to the relativistic
classical electrodynamics is missing as yet.

\subsection*{Acknowledgements} The author is indebted to Petr Jizba,
Ji\v{r}\'{\i} Tolar and Uwe-Jens Wiese for useful discussions.

\end{document}